\documentstyle[12pt,epsf]{article}

\begin{document}

\begin{titlepage}

\hfill PITHA 94/9

\hfill hep-lat/9401034

\begin{centering}
\vfill

{\bf THE UNIVERSAL EFFECTIVE POTENTIAL \\
FOR THREE-DIMENSIONAL MASSIVE SCALAR FIELD THEORY \\
FROM THE MONTE CARLO STUDY OF THE ISING MODEL
}

\vspace{1cm}

M.M.Tsypin\footnote{Alexander von Humboldt Fellow

\ \ \ e-mail: hkf256@djukfa11.bitnet}

\vspace{1cm}

{\em Institute for Theoretical Physics E,
RWTH Aachen, D-52056 Aachen, Germany}

\vspace{0.3cm}

and

\vspace{0.3cm}

{\em Department of Theoretical Physics, \\
P.~N.~Lebedev Physics Institute, Leninsky pr.~53, 117924 Moscow, Russia}

\vspace{2cm}
{\bf Abstract}

\end{centering}

\vspace{0.3cm}\noindent

We study the low-energy effective action $S_{eff}[\varphi]$
for the one-component real scalar field theory in three
Euclidean dimensions in the symmetric phase, concentrating
on its static part --- effective potential $V_{eff}(\varphi)$.
It characterizes the approach to the phase transition in all
systems that belong to the 3d Ising universality class.
We compute it from the probability distributions of the
average magnetization in the 3d Ising model in a homogeneous
external field, obtained by Monte Carlo. We find that the
$\varphi^6$ term in $V_{eff}$ is important, while the higher
terms can be neglected within our statistical errors.
Thus we obtain the approximate effective action
\begin{displaymath}
  S_{eff} = \int d^3 x
  \left\{
  {1 \over 2} \partial_\mu \varphi \partial_\mu \varphi +
  {1 \over 2} m^2 \varphi^2 + m g_4 \varphi^4 + g_6 \varphi^6
  \right\} ,
\end{displaymath}
with arbitrary mass $m$ that sets the scale, and dimensionless
couplings $g_4 = 0.97 \pm 0.02$ and $g_6 = 2.05 \pm 0.15$.
The value of $g_4$ is consistent with the renormalization group
fixed point coupling. This $V_{eff}$, when used instead of
the traditional $a \varphi^2 + b \varphi^4$, turns the
Ginzburg--Landau description of the long-wave properties
of the 3d theory near criticality into quantitatively accurate.
It is also relevant to the theory of cosmological
phase transitions.

\vfill \vfill

\noindent
PITHA 94/9

\noindent
January 1994

\end{titlepage}

\section{Introduction}
This work is devoted to the following problem: what is the effective
potential, and the corresponding effective Ginzburg--Landau theory,
that would provide not exact, but reasonably phenomenologically accurate
description of the properties of the 3d Ising model near the phase
transition (and other models that belong to the same universality class)?

The model from this universality class that is particularly
suitable for field-the\-or\-eti\-cal treatment is the theory of
one-component real scalar field in three Euclidean dimensions
(``3d $\phi^4$ theory''), defined by the (bare) action
\begin{equation} \label{phi4}
  S = \int d^3 x
  \left\{
  {1 \over 2} \partial_\mu \phi \partial_\mu \phi +
  {1 \over 2} m^2 \phi^2 + \lambda \phi^4
  \right\} .
\end{equation}
Thus, from the field-theoretical point of view, we study the
low-energy effective action of this theory.

This problem, being interesting by itself, is also relevant
to the theory of cosmological phase transitions in the early
Universe. The second order high-temperature phase transition
in the 3+1-dimensional quantum field theory is in the
universality class of the 3d Euclidean phase transition.
The weak first-order high-temperature transitions can be studied
in the framework of effective 3d Euclidean theory as well.
The effective potential for such problems has been a subject
of recent investigations~\cite{KaRu93,Sh93}. The use of the
perturbation theory is hindered in three dimensions
by infrared divergences and by the strong-coupling nature
of the problem. Such issues as the existence and role of the
$|\varphi|^3$ term in the effective potential remain to be
settled.

Thus the nonperturbative study of the effective action of the
simplest 3d field theory (\ref{phi4}), or that of the 3d
Ising model, seems appropriate.

\section{The model}
We study the Ising model with the nearest-neighbour interaction
on a simple cubic lattice. The partition function is
\begin{equation} \label{Ising}
  Z = \sum_{ \{\phi_i \} } \exp \Bigl\{
  \beta \sum_{<ij>} \phi_i \phi_j + J \sum_i \phi_i \Bigr\},
  \qquad \phi_i = \pm 1,
\end{equation}
where $J$ is the homogeneous external field (``magnetic field'').
We study the symmetric (paramagnetic) phase, thus the coupling
$\beta$ is less than, but close to the critical value
$\beta_c \approx 0.22165$.

Our main subject are the long-wave (low-momentum, low-energy)
properties of the model, when it is in the scaling region, but
not exactly at the critical point. Then the properties are fixed,
the only free parameter is the mass ( = the scale). The particles
of the corresponding 2+1-dimensional field theory are massive
(and thus can be nonrelativistic) and have well-defined low-energy
properties, such as nonrelativistic scattering amplitudes.
The effective action we are looking for is a convenient formalism
to describe these properties.

\section{The effective action}
The low-energy Ginzburg-Landau-Wilson effective action can be
written as
\begin{equation} \label{Seff}
  S_{eff} = \int d^3 x \left\{
  {1 \over 2} Z_\varphi^{-1} \partial_\mu \varphi \partial_\mu \varphi
  + V_{eff}(\varphi) - J(x) \varphi(x)
  \right\},
\end{equation}
where $\varphi(x)$ is the (slowly varying) average magnetization,
and we keep only the lowest-order derivative term. To compute
$S_{eff}$ one needs to know the effective potential $V_{eff}(\varphi)$
and the field renormalization factor $Z_\varphi$. To compute
the former, it is sufficient to consider only the homogeneous
external field $J(x)=J$; the latter can be derived from the
two-point correlation function of $\varphi$. Thus we never have
to work with the explicitly $x$-dependent $J(x)$.

\subsection{The effective potential}
The effective potential (free energy)
$V_{eff}(\varphi)$~\cite{Goldstone}-\cite{CW} is defined by
\begin{equation} \label{Veff}
 { d V_{eff}(\varphi) \over d \varphi } = J,
 \qquad \varphi = \langle \phi \rangle_J.
\end{equation}
Thus~(\ref{Seff}), considered at the tree level, reproduces correctly
the average magnetization $\langle \phi \rangle_J$ as a function of
external field $J$. We recall also that $V_{eff}(\varphi)$ is a
generating function for one-particle irreducible (1PI) $n$-point
Green functions $\Gamma_n$ at momentum zero,
\begin{equation}
  V_{eff}(\varphi) = \sum_n {1 \over n!} \Gamma_n(p=0) \varphi^n .
\end{equation}

\subsection{How to compute $V_{eff}$ on the lattice -
existing methods}
Our goal is to compute $V_{eff}$ by Monte Carlo with the best
possible precision. There are at least three approaches.

1) Direct computation of the correlation functions $\Gamma_n$,
such as
\begin{equation}
 \Gamma_4(p=0) = \langle \phi^4 \rangle - 3 \langle \phi^2 \rangle^2,
\end{equation}
and so on~\cite{FrSm82}-\cite{Wh84}. This approach works reasonably well
for $\Gamma_4$, but is hardly feasible for $\Gamma_6$ and higher
$\Gamma_n$: after subtraction of disconnected and one-particle
reducible parts the signal to noise ratio turns out to be very
bad, and statistical errors are prohibitively high~\cite{Wh84}.

2) Compute the magnetization per spin $\varphi= \langle \phi \rangle_J$
as a function of the external field $J$ by Monte Carlo, invert this
function to obtain $J=J(\varphi)$ and integrate it numerically
to obtain $V_{eff}(\varphi)$ according
to~(\ref{Veff})~\cite{CaMa83,Ca83}.
The drawback of this approach is that it requires many
measurements of $\langle \phi \rangle_J$ at different values of $J$.

3) Study the probability distribution of the order parameter
${1 \over N} \sum_i \phi_i$~\cite{Bi81}. One uses the Monte Carlo
algorithm to generate a Boltzmann ensemble of configurations.
For every configuration in the ensemble one computes the
order parameter (magnetization per site)
$\varphi = {1 \over N} \sum_i \phi_i$, where $N$ is the total
number of sites on the lattice. Thus the probability distribution
of $\varphi$ is obtained, which is characterized by the probability
density
\begin{equation}
  P(\varphi) = {1 \over Z}
  \sum_{ \{\phi_i \} }
  \delta \Bigl( {1 \over N} \sum_i \phi_i - \varphi \Bigr)
  \exp \Bigl\{\beta \sum_{<ij>} \phi_i \phi_j  \Bigr\}.
\end{equation}
To prove that this is indeed the probability density, it is sufficient
to observe that for any function $f(\varphi)$
\begin{equation}
  \int P(\varphi) f(\varphi) d \varphi =
  \Bigl\langle f \Bigl( {1 \over N} \sum_i \phi_i \Bigr) \Bigr\rangle .
\end{equation}
This probability is considered usually in connection with the
``constrained effective potential'' $V_{con}(\varphi)$
\cite{Goldstone},\cite{Bi81}-\cite{GoLe91}, which is defined by
\begin{equation} \label{Vcon}
  P(\varphi) \propto \exp \left\{ - \Omega V_{con}(\varphi) \right\},
\end{equation}
for a system in a box of volume $\Omega$ (we set the lattice
spacing equal to unity, so $\Omega = N$). It can be shown that
in the infinite volume limit the constrained effective potential
coincides with the standard $V_{eff}$~\cite{GiLe83,RaWi86}.

\begin{figure}
\epsfbox{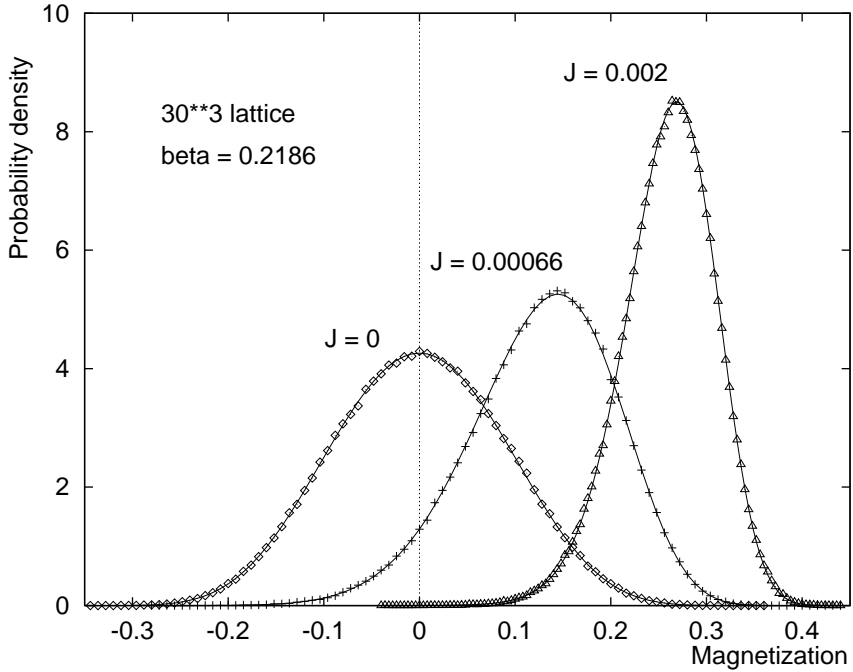}
\caption{
A set of probability densities $P(\varphi)$ for the
magnetization per lattice site $\varphi$, for the Ising model
(\protect\ref{Ising}). The solid line is the fit
with~(\protect\ref{preexp}),
$V_{eff}(\varphi) = r \varphi^2 + u \varphi^4 + w \varphi^6$.
Three histograms are fitted simultaneously with the same $V_{eff}$.
 }
\label{fig1}
\end{figure}

One can gain additional insight into these different ways to
extract $V_{eff}$ from the Monte Carlo data by considering
the actual distributions of the order parameter in a typical
case (Fig.~\ref{fig1}).

The first method studies, essentially, the deviation of the
$J=0$ curve from the Gaussian, which is quite small and difficult
to measure. The distribution is located at small values of $\varphi$,
where the dominating term in $V_{eff}$ is $\varphi^2$, while
higher terms provide only small corrections.

The second method means that we use each curve to compute
the expectation value of $\varphi$, i.e. $\langle \phi \rangle_J$,
and discard all remaining information contained in the data.
(It seems that one can obtain a good method by combining this with
the reweighting technique~\cite{FeSw88,FeSw89}. Then
$\langle \phi \rangle_J$ could be computed accurately for
any $J$ from 0 to $\approx 0.3$ just from the data at
Fig.~\ref{fig1} plus the data on the energy. And all the
information from probability distributions  is utilized.
However, we do not pursue this line further here).

The third method also looks at the form of the $J=0$ probability
distribution and thus has the similar drawbacks as the first one.

\subsection{Our method}
For the Monte Carlo computation of the effective potential
we have developed a method that is close in spirit to the
constrained effective potential approach, but contains
two significant improvements.

First, it is obvious from the preceding discussion, that
to obtain the information on the higher powers of $\varphi$
in $V_{eff}$, one should study the system at $\varphi$ far
from zero. This can be achieved with the help of the
external field. The definition~(\ref{Vcon}) allows a
straightforward generalization for nonzero external field:
\begin{equation} \label{eq14}
  P(\varphi) \propto \exp \left\{
  - \Omega V_{con}(\varphi) + \Omega J \varphi \right\}.
\end{equation}
Thus one can check whether some ansatz, such as
$V_{con}(\varphi) = r \varphi^2 + u \varphi^4$, gives a good
approximation for the $V_{con}$, and find the values of
parameters (such as $r$ and $u$), by performing a simultaneous
fit of several histograms, corresponding to different values
of $J$, by~(\ref{eq14}), using the parameters of $V_{con}$
as fit parameters.

\begin{figure}
\epsfxsize=12.5cm\epsfbox{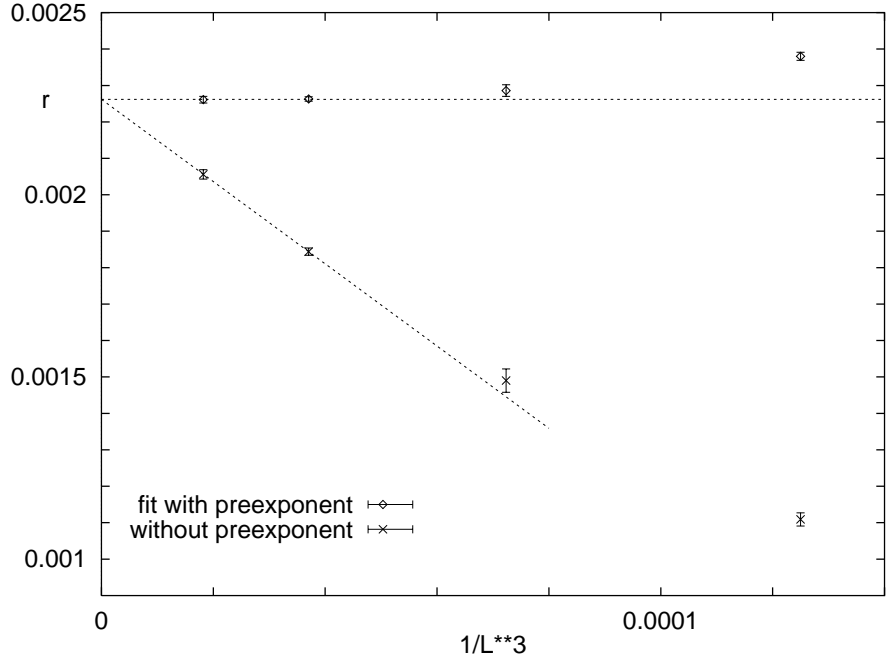}
\caption{
Finite volume dependence of the parameter $r$, obtained by fitting
the probability distributions for $\beta = 0.2186$, $J=0$, 0.00066
and 0.002 --- the same values as at Fig.~\protect\ref{fig1} --- with
(\protect\ref{nopreexp}) and (\protect\ref{preexp}),
$V_{eff}(\varphi) = r \varphi^2 + u \varphi^4 + w \varphi^6$,
for different lattice sizes: $20^3$, $24^3$, $30^3$ and $38^3$
(the correlation length $\xi \approx 7.4$). Parameters $u$ and $w$
are less volume-sensitive.
 }
\label{fig2}
\end{figure}

The second point concerns the correspondence between the probability
distribution and the effective potential. When we use the
relation~(\ref{eq14}), we get the finite-volume constrained
effective potential. However, the quantity that enters the
low-energy effective action is the {\em infinite}-volume $V_{eff}$,
i.e., the free energy per unit volume in the usual thermodynamical
sense. As in the infinite-volume limit ($\Omega \to \infty$)
both potentials are known to coincide, one is tempted to use
the formula~\cite{ChLa86}
\begin{equation} \label{nopreexp}
  P(\varphi) \propto \exp \left\{
  - \Omega V_{eff}(\varphi) + \Omega J \varphi \right\}.
\end{equation}
and to hope that for the reasonably large volume $\Omega$
the finite volume effects are reasonably small. It turns out
that {\em this is actually not the case}. The situation is illustrated
by the lower graph at the Fig.~\ref{fig2}. The finite volume
corrections (which manifest themselves by the volume dependence
of parameters of $V_{eff}$) are unacceptably large, in spite
of going to zero in the infinite volume limit. This behaviour
shows that~(\ref{nopreexp}) is not a very good approximation.
It turns out that while~(\ref{nopreexp}) reproduces correctly
the exponential dependence on $\Omega$ at $\Omega \to \infty$,
it misses the important preexponential factor.

The improved formula for the probability distribution
of the order parameter of the system in a finite box
of volume $\Omega$ with the periodic boundary conditions is
\begin{equation} \label{preexp}
  P(\varphi) \propto
  \sqrt{ d^2 V_{eff}(\varphi) \over d \varphi^2 }
  \exp \left\{   - \Omega V_{eff}(\varphi) + \Omega J \varphi \right\},
\end{equation}
where $V_{eff}(\varphi)$ is the infinite-volume effective potential.
This relation can be found (in somewhat implicit form or for
special cases) in the literature (see~\cite{BrZi85}; \cite{Pr90},(5.54);
\cite{GoLe91},(4.27)). The upper graph at Fig.~\ref{fig2}
confirms that this is indeed a good approximation: no finite
volume corrections are visible for $L/\xi \ge 4$.

We outline here briefly the derivation of the preexponential factor
in~(\ref{preexp}). On the one-loop level one can write the effective
potential as a bare potential plus one-loop correction. For a
theory with a bare Lagrangian
\begin{equation}
 {\cal L}_0 = {1 \over 2} \partial_\mu \phi \partial_\mu \phi +
 V_0(\phi)
\end{equation}
we have
\begin{equation} \label{eq17a}
  V_{eff}(\varphi) = V_0(\varphi) +
  {1 \over 2} \int {d^3 p \over (2 \pi)^3}
  \ln \Bigl( p^2 + {d^2 V_0 \over d \varphi^2} \Bigr).
\end{equation}
At the same time for the constrained effective potential
in the box of volume $\Omega$ the integral is substituted
by a sum over momenta allowed by the boundary
conditions, with the $p=0$ mode excluded:
\begin{equation} \label{eq17b}
  {V_{con}(\varphi)|}_\Omega = V_0(\varphi) +
  {1 \over 2 \Omega}
  {\sum_p}'
  \ln \Bigl( p^2 + {d^2 V_0 \over d \varphi^2} \Bigr).
\end{equation}
Thus the difference between $V_{con}$ and $V_{eff}$ reduces
to the difference between the sum and the integral. This difference
has been computed in~\cite{BrZi85}, sect.~3.3. The leading large
volume term takes a simple form:
\begin{equation}
  {V_{con}(\varphi)|}_\Omega - V_{eff} =
  - {1 \over 2 \Omega}
  \ln  {d^2 V_0 \over d \varphi^2} .
\end{equation}
This provides the preexponential factor in~(\ref{preexp}).
The last step is to substitute $V_0$ by $V_{eff}$, which
means essentially the use of selfconsistent values for the masses.

\subsection{
The preexponent in the probability distribution \protect\\
and the equal weight versus equal height problem \protect\\
for asymmetric first order phase transitions}
We deviate a little from our problem to make an additional comment
on the formula~(\ref{preexp}).

This formula seems quite universal and useful for the
various problems connected with the order parameter probability
distribution. In our problem we use it in the symmetric phase
for a range of $J$. A very different and in some sense complementary
application is the equal weight versus equal height problem for
asymmetric first order transitions~\cite{ChLa86},\cite{Pr90}-\cite{PrRu90}.
Consider a model with a first order phase transition between
two phases characterized by the order parameters $\varphi_1$
and $\varphi_2$, exactly at the transition point. This means
that the free energies for both phases are the same, i.e.,
$V_{eff}(\varphi_1) = V_{eff}(\varphi_2)$. Put the system
in a finite box much larger than the correlation lengths
in both phases. Consider the probability distribution $P(\varphi)$
of the order parameter. Then this distribution can be approximated
by the two narrow Gaussian peaks around $\varphi_1$ and
$\varphi_2$~\cite{BiLa84,ChLa86}. The question is, do they
have equal height or equal weight? The formula~(\ref{nopreexp})
predicts equal height, while it can be shown that equal weight
is the correct answer~\cite{BoKo90}-\cite{PrRu90},\cite{Pr90}. But this
it just what follows from~(\ref{preexp}).

\section{Monte Carlo computation of the effective action}
\begin{figure}
\epsfbox{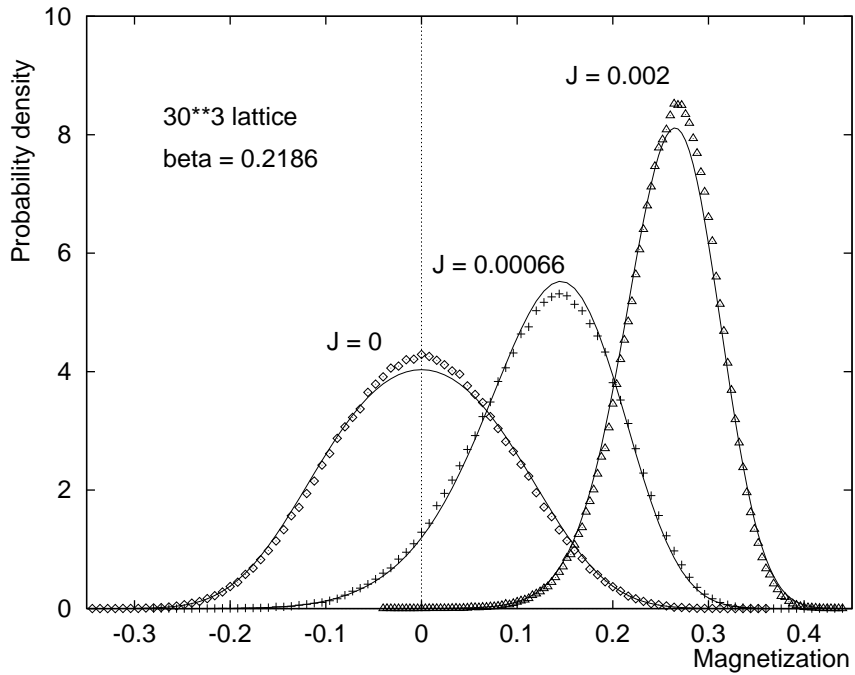}
\caption{
The data is the same as at Fig.~\protect\ref{fig1}. The solid line is the
best simultaneous fit with (\protect\ref{preexp}),
$V_{eff}(\varphi) = r \varphi^2 + u \varphi^4$.
Statistics: 400000 configurations for every histogram.
 }
\label{fig3}

%\vspace*{10cm}

\end{figure}

Now we turn to our computation of the effective action.
We study the 3d Ising model~(\ref{Ising}) on a simple cubic lattice
with periodic boundary conditions, on lattices from $14^3$ to
$58^3$. The Swendsen-Wang cluster Monte Carlo algorithm in the external
magnetic field~\cite{SwWa87}-\cite{Wa89} is used to generate the
Boltzmann ensemble of configurations. (We use the version of this
algorithm without the ghost spin). For every configuration we
measure magnetization per site $\varphi = {1 \over N} \sum_i \phi_i$
and compute the histograms for the probability density $P(\varphi)$,
for several values of $J$. Then we do the simultaneous fit of
all the histograms with the formula~(\ref{preexp}). (We minimize the sum
of $\chi^2$ from the individual histograms). The first ansatz to
try is
\begin{equation} \label{v4}
  V_{eff}(\varphi) = r \varphi^2 + u \varphi^4,
\end{equation}
inspired by the standard Ginzburg-Landau theory (or the tree-level
$\phi^4$ theory). The result is shown at Fig.~\ref{fig3}.
One can see the discrepancy between the data and the fit ---
not very large, but statistically significant at our level
of precision. This means that~(\ref{v4}) does not provide a good
quantitative description of the effective potential. So we
consider a three-parameter expression
\begin{equation} \label{v6}
  V_{eff}(\varphi) = r \varphi^2 + u \varphi^4 + w \varphi^6.
\end{equation}
This form is motivated by several intuitive considerations that
are discussed below. We have found that it provides the ideal
fit (Fig.~\ref{fig1}): there is no systematic discrepancy
between the data and the fit, just statistical noise.
We have found no other reasonable ansatz that works so well
(we have also tried three-parameter expressions
$V_{eff}(\varphi)= r \varphi^2 + u |\varphi|^w$ and
$V_{eff}(\varphi)= r \varphi^2 + w|\varphi|^3 + u \varphi^4$
--- they work poorly).

Thus for every value of the bare coupling $\beta$ we obtain
the low-energy effective Lagrangian
\begin{equation}
  {\cal L}_{eff} = {1 \over 2} Z_\varphi^{-1}
  \partial_\mu \varphi \partial_\mu \varphi +
  r \varphi^2 + u \varphi^4 + w \varphi^6.
\end{equation}
The three parameters $r$, $u$, $w$ are determined by the fitting
procedure described above. The field renormalization factor $Z_\varphi$
is obtained from the propagator
\begin{equation}
  G_2( {\bf p} ) = \langle \phi({\bf p})
  \phi^*({\bf p}) \rangle,
\end{equation}
where
\begin{equation}
  \phi({\bf p}) = {1 \over \sqrt{N} } \sum_{\bf x}
  \phi_{\bf x} e^{ i{\bf px}}.
\end{equation}
Then at small momentum $p$
\begin{equation}
  G_2( {\bf p} )^{-1} = Z_\varphi^{-1} p^2 + 2 r.
\end{equation}
We use the lattice version of this,
\begin{equation}
  G_2( {\bf p} )^{-1} = Z_\varphi^{-1}
  \sum_{\mu=1}^3 2(1 - \cos p_\mu) + 2 r,
\end{equation}
and use typically three values of momentum,
\begin{equation}
  p_\mu = (p,0,0), \quad p = { 2\pi n \over L}, \quad n=0,1,2,
\end{equation}
to find $Z_\varphi^{-1}$ as the slope of $ G_2( {\bf p} )^{-1} $
as a function of $\sum_\mu 2(1-\cos p_\mu)$.

After the renormalization of $\varphi$,
\begin{equation}
 \varphi = \sqrt{Z_\varphi} \varphi_R,
\end{equation}
we obtain the effective Lagrangian in the form
\begin{equation} \label{eq22}
  {\cal L}_{eff} =
  {1 \over 2} \partial_\mu \varphi_R \partial_\mu \varphi_R +
  {1 \over 2} m^2 \varphi_R^2 + m g_4 \varphi_R^4 + g_6 \varphi_R^6,
\end{equation}
where
\begin{equation}
  m = \sqrt{2 Z_\varphi r}, \quad
  g_4 = {Z_\varphi^2 u \over \sqrt{2 Z_\varphi r} }, \quad
  g_6 = Z_\varphi^3 w.
\end{equation}
In the continuum limit ($m \to 0$) this effective Lagrangian
should be universal. Thus the only free parameter is $m$,
that determines the scale, while the dimensionless four- and
six-point couplings $g_4$ and $g_6$ take definite values
that are the same for the whole 3d Ising universality class.
Our numerical results are
collected in Table~\ref{table1} and
represented at Figs.~\ref{fig4}--\ref{fig7}.

\section{Data analysis and extrapolation \protect\\
to the continuum limit}
\begin{figure}
\epsfbox{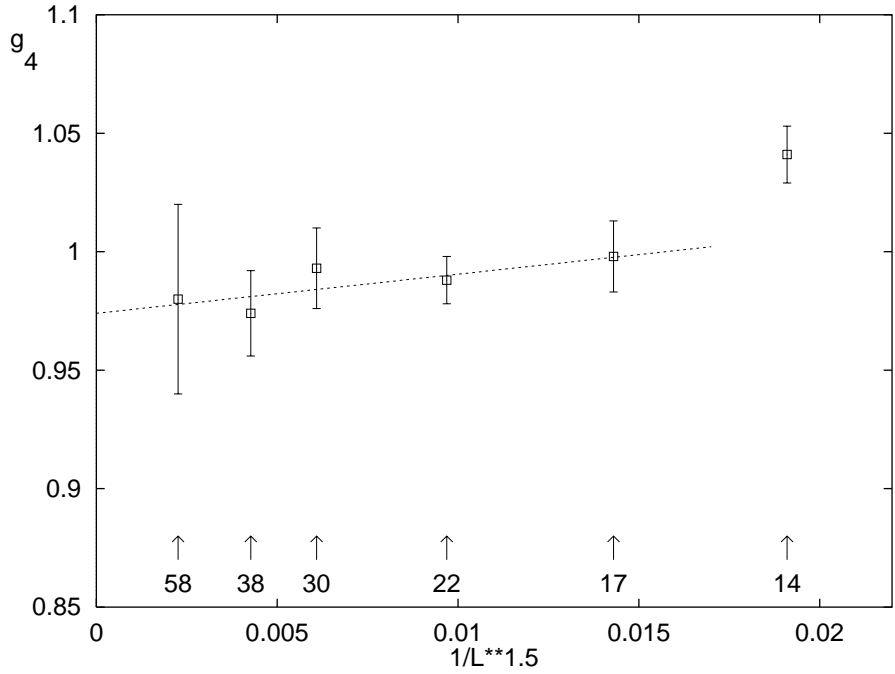}
\caption{
The dimensionless four-point coupling $g_4$ as a function of the
lattice size $L$. The ratio $L/\xi$ is kept fixed at around 4.1.
The errors shown are standard deviations.
 }
\label{fig4}
\end{figure}

\begin{figure}
\epsfbox{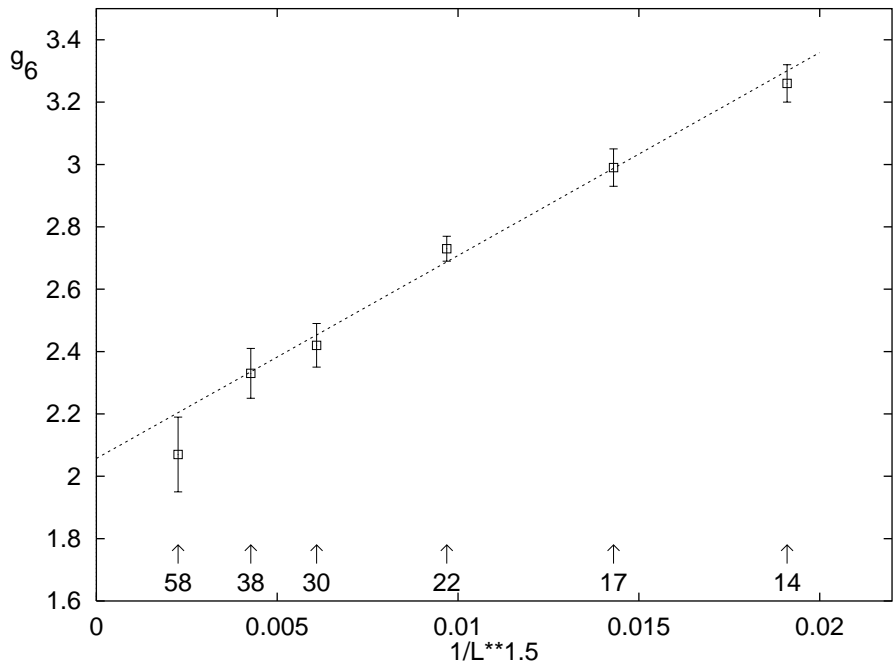}
\caption{
The six-point coupling $g_6$ as a function of lattice size $L$.
($L/\xi \approx 4.1$).
 }
\label{fig5}
\end{figure}

\begin{figure}
\epsfbox{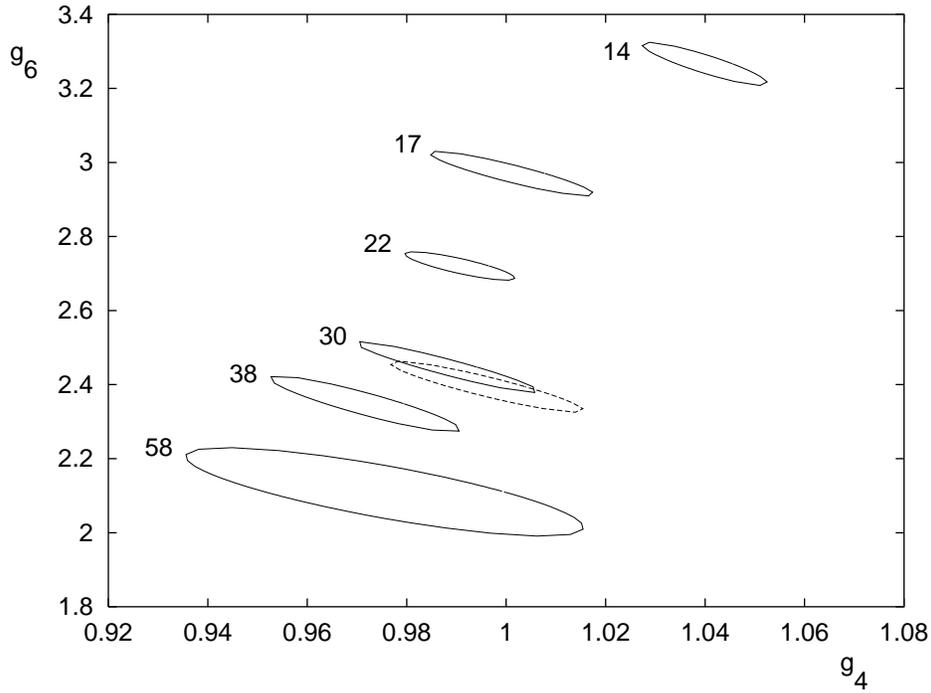}
\caption{
The simultaneous plot of $g_4$ and $g_6$. The ellipses show the
statistical errors (semiaxis = standard deviation). The lattice size
$L$ is shown near the ellipses, $L/\xi \approx 4.1$.
The broken line is for the same parameters ($\beta$ and three values
of $J$) as for $L = 30$, but measured on a $38^3$ lattice.
This provides the additional check for the smallness of the
finite volume corrections.
 }
\label{fig6}
\end{figure}

\begin{figure}
\epsfbox{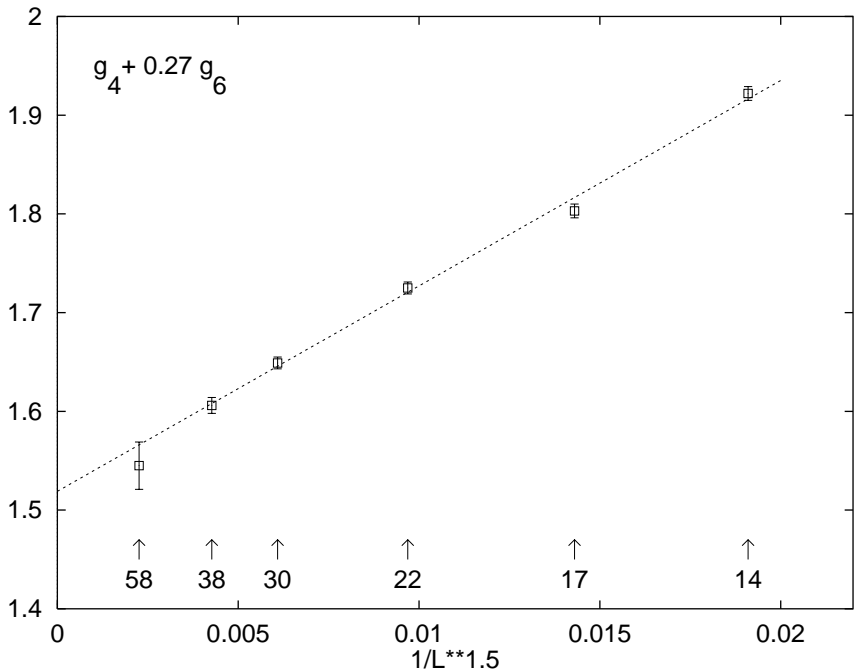}
\caption{
The linear combination of dimensionless couplings that is determined
with the smallest statistical error.
 }
\label{fig7}
\end{figure}

Apart from statistical errors, there are two sources of
systematic errors: finite volume and finite UV cutoff.
To check for the finite volume effects, we increase the lattice
size $L$, keeping $m$ fixed. One can see from Fig.~\ref{fig2}
that for $L/\xi \ge 4$ finite volume effects are negligible.
To check for the effect of the finite UV cutoff, we keep
$L/\xi$ fixed at $\approx 4.1$, increase $\xi$ and scale $J$
according to
\begin{equation}
  J \propto \xi^{-\beta \delta / \nu}
  \qquad (\beta \delta \approx 1.57,\ \nu \approx 0.63)
\end{equation}
The scaling limit is characterized by stabilization
of $g_4$ and $g_6$ as the functions of $\xi$, for
large enough $\xi$. We observe a smooth approach to scaling
(Figs.~\ref{fig4} and \ref{fig5}), but note that the shift of
$g_6$ caused by the finiteness of the cutoff is still visible
on our largest lattices. The correlation length
in zero field is $\xi = m^{-1} \approx 14$ for the $58^3$ lattice,
but one should keep in mind that it is smaller for nonzero
$J$, and for the largest $J$ used in the fit it is approximately
twice as small. The presence of finite cutoff effects in our data
makes it necessary to extrapolate the results to continuum
limit $\xi \to \infty$. As we keep $L/\xi$ fixed, we
work in terms of $L$. The reasonable extrapolation is
\begin{equation}
  g_6(L) = g_6(\infty) + a L^{-\kappa}.
\end{equation}
The statistical errors in the data for $g_6$ are still
too high to allow the determination of the exponent $\kappa$
with reasonable accuracy, while different values of $\kappa$
(such as $\kappa=1$ or $\kappa=2$) lead to considerably
different extrapolated values $g_6(\infty)$. The situation can
be alleviated if we take into account that the statistical
errors of $g_4$ and $g_6$ are not independent (Fig.~\ref{fig6}),
and so there is a linear combination of $g_4$ and $g_6$
that has much smaller error than $g_4$ and $g_6$ separately
(Fig.~\ref{fig7}). For this combination one can find
the corresponding exponent $\kappa = 1.5 \pm 0.2$. The plausible
assumption is that $g_6$ should be extrapolated with the
same exponent ($g_4$ shows little dependence on $L$ for
$ L \ge 17$, so one does not really have to extrapolate it).
That is why we plot the couplings at Figs.~\ref{fig4} and
\ref{fig5} as functions of $L^{-1.5}$.

Our result for the continuum limit is
\begin{equation} \label{result}
\begin{array}{l}
  g_4 = 0.97 \pm 0.02, \\
  g_6 = 2.05 \pm 0.15,
\end{array}
\end{equation}
where the errors are the standard deviations.

\section{Discussion}
Here we compare our results with the data available in the
literature, and give some semiintuitive arguments in favour
of the effective Lagrangian (\ref{eq22}) as a good approximate theory.

\subsection{Four-point coupling}
The four-point coupling of the 3d scalar field theory has been
a subject of many studies. It is not the main subject of our
investigation (our main point is the role of $g_6$), but
the comparison of our $g_4$ with the available data provides
a useful consistency check of our computation. Here we list some data
for $g_4$ obtained by different methods.

\medskip

\noindent
{}From the 3d renormalization group fixed point:

\smallskip

\begin{tabular}{ll}
$0.989 \pm 0.004$ & \cite{BaNi76}-\cite{LeZi80} \\
0.981             & \cite{Ni91}, revised estimate
\end{tabular}

\smallskip

\noindent
{}From the high-temperature series~\cite{BaKi79}-\cite{Ba84}:

\smallskip

\begin{tabular}{ll}
$0.99 \pm 0.03$    & (simple cubic lattice) \\
$0.991 \pm 0.003$  & (BCC lattice) \\
$0.99 \pm 0.01$    & (FCC lattice)
\end{tabular}

\smallskip

\noindent
{}From the Monte Carlo studies of the
$\langle\phi^4\rangle - 3 \langle\phi^2\rangle^2$:

\smallskip

\begin{tabular}{ll}
$1.00 \pm 0.04$  & \cite{FrSm82}, $\xi = 3.3$ \\
$1.00 \pm 0.03$  & \cite{FrBa82}, $\xi = 6.6$ \\
$0.80 \pm 0.02$  & \cite{FrBa82}, $\xi = 16$  \\
$0.95 \pm 0.08$  & \cite{Wh84}, $\xi = 3.2$ \\
$0.90 \pm 0.04$  & \cite{We89}, extrapolation to the continuum limit \\
$1.00 \pm 0.08$  & \cite{KiPa93}, $\xi = 14.5$
\end{tabular}

\smallskip

\noindent
We observe that our result $g_4 = 0.97 \pm 0.02$ fits well into
this picture.

\subsection{Six-point coupling}
The information about $g_6$ is more scarce. The only Monte Carlo
study we are aware of is~\cite{Wh84}. However, large statistical
errors made it impossible to reach a definite conclusion
about the value of $g_6$ in the continuum limit and whether
it is different from zero.

Another source of information is provided by the study of the
``Ising equation of state'' in the framework of the
$\varepsilon$-expansion~\cite{AvMi72}-\cite{Zi89}.
The equation of state describes magnetization as a
function of the homogeneous external field, thus providing
information on the effective potential. As the gradient
term in ${\cal L}_{eff}$, that determines the normalization
of the renormalized field, is not considered, one can
extract parameters that do not depend on this normalization.
The ratio $g_6/g_4^2$ is such a parameter invariant
under the change of the scale of the field. The convenient
representation of the equation of state is given by
Avdeeva and Migdal~\cite{AvMi72}. In original notation,
\begin{equation}
  H = \chi^{-(\beta + \gamma)/\gamma} \phi(M \chi^{\beta/\gamma}),
\end{equation}
where $M$ is magnetization, $H$ is the external field,
$\chi = \partial M / \partial H$ is the susceptibility
in the finite field, and the function $\phi$ is found to be
\begin{equation}
  \phi(m) = m - m^3 + {\varepsilon^2 \over 4} m^5 + {\cal O}(\varepsilon^3).
\end{equation}
for the dimension of space $d=4-\varepsilon$. From this equation
it is straightforward to compute
\begin{equation}
   {g_6 \over (g_4)^2} = 2 \varepsilon - {20 \over 27} \varepsilon^2
                         + {\cal O}(\varepsilon^3).
\end{equation}
The next order in $\varepsilon$ can be derived from the parametric
representation of the equation of state described in~\cite{Zi89},
sect. 25.1.3. One finds
\begin{equation}
   {g_6 \over (g_4)^2} = 2 \varepsilon - {20 \over 27} \varepsilon^2
    + 1.2759 \varepsilon^3 + {\cal O}(\varepsilon^4).
\end{equation}
The series is obviously divergent, and the only straightforward
conclusion for $\varepsilon = 1$ that one can make
without some resummation is
\begin{equation}
   {g_6 \over (g_4)^2} = 2.0 \pm 0.7 .
\end{equation}
This agrees well with our result~(\ref{result}).

The effective potential of the $\phi^4$ model can be also
computed in the framework of perturbation theory directly
in $d=3$. The result including up to five loops can be found
in~\cite{HaDo92}, eq.(3.3). It turns out, however, that
the two-loop contribution to $g_6$ is four times as large as
the one-loop, the three-loop even larger, etc. This makes
it impossible to derive any number for $g_6$ without some
resummation of the series.

Yet another approach to the computation of the effective Lagrangian
is provided by the Wegner-Houghton equation in the local-potential
approximation~\cite{HaHa86,BaBe90}. The fixed point for the
potential corresponds to $g_6 = 2.40$~\cite{BaBe90}.

One more approach is the strong-coupling
expansion~\cite{BeCo80}-\cite{BeCo81b} and the dimensional
expansion~\cite{BeBo93} for the field theory in the Ising
limit, i.e., in the limit of infinitely strong bare coupling.
The results obtained for $d=3$ disagree with ours, as the
strong-coupling expansion favours $g_6=0$~\cite{BeCo81b},
while the dimensional expansion leads to conjecture that
$g_6 = \infty$~\cite{BeBo93}.

Finally, we compare our $g_6$ with results of Tetradis
and Wetterich, obtained within the ``effective average action''
approach~\cite{ReTe93}-\cite{TeWe93b}. They use the symbol
$u_3$ for the six-point coupling, the correspondence is
$g_6 = u_3/48$. The fixed-point value is
$u_{3*} = 87.4$ (\cite{TeWe93a}, Table 2), while
the asymptotic value of the low-energy coupling,
as the phase transition is approached from the symmetric phase
(which is more appropriate to compare with our $g_6$),
is $u_{3S} = 107$~\cite{Te93}.
This corresponds, respectively, to $g_6=1.82$ and $g_6 = 2.23$.

\subsection{Philosophy}
A widespread point of view on the effective potential in 3d
is as follows. The problem should be considered in the
framework of the $\phi^4$ theory. Then either you work
on the tree level, and have a standard Landau theory with
\begin{equation} \label{phil4}
  V_{eff}(\varphi) = a \varphi^2 + b \varphi^4,
\end{equation}
or you include loop corrections, and then you must retain all
powers of $\varphi$ in $V_{eff}$. The $\varphi^6$ term
should be considered on equal footing with other higher terms.

Our study corroborates an alternative point of view advocated
by Tetradis and Wetterich~\cite{TeWe93a,TeWe93b}: that
while~(\ref{phil4}) is a rather rough approximation, the
ansatz
\begin{equation} \label{phil6}
  V_{eff}(\varphi) = {1 \over 2} m^2 \varphi^2 + m g_4 \varphi^4 +
  g_6 \varphi^6
\end{equation}
gives a very good approximation, and the higher powers of
$\varphi$ can be considered as small corrections. Additional
arguments in favour of this point of view can be drawn
from the smallness of the critical index $\eta$ in the 3d theory.

Consider the asymptotic behaviour of magnetization
$\varphi=\langle\phi\rangle_J$ in the external field as
a function of $J$ at large $\varphi$ ( = large $J$).
This is the same limit as if we keep $J$ fixed and let $m \to 0$.
The equation $dV_{eff}/d\varphi = J$ gives
\begin{equation}
 6 g_6 \varphi^5 = J, \quad \langle\phi\rangle_J \propto J^{1/5}.
\end{equation}
The Landau theory~(\ref{phil4}) at the critical point ($a=0$)
would give
\begin{equation}
  \varphi^3 \propto J, \quad \langle\phi\rangle_J \propto J^{1/3}.
\end{equation}
However, this is exactly the definition of the critical index $\delta$:
\begin{equation}
  \langle\phi\rangle_J \propto J^{1/\delta}.
\end{equation}
Thus the Landau theory~(\ref{phil4}) corresponds to $\delta=3$,
and~(\ref{phil6}) corresponds to $\delta=5$, while the correct
value is
\begin{equation}
  \delta = {d+2-\eta \over d-2+\eta} \approx 4.8 .
\end{equation}
We see that while $\delta \ne 5$ means that (\ref{phil6})
is, strictly speaking, inapplicable for very large $\varphi$
(such that $V_{eff}''(\varphi) \gg V_{eff}''(0) = m^2$),
5 is still a much better approximation to $\delta$ than 3,
due to the smallness of $\eta$. In the large $N$ limit of
the $O(N)$ Heisenberg model in 3d $\eta \to 0$, and the
$N$-component analog of (\ref{phil6}) becomes
exact~\cite{ReTe93,El93}.

The form~(\ref{phil6}) is inapplicable for $m=0$, i.e. exactly
at the transition point. In this case every nonzero $\varphi$
is `very large', and the correct form is
\begin{equation}
  V_{eff} \propto \varphi^{\delta+1} \qquad (m=0).
\end{equation}
However, the whole idea of the low-energy effective action is no
more applicable in this case, because `low energy' means $p \ll m$.
So we always work at nonzero $m$.

Notice that $\varphi^6$ is renormalizable in 3d, while higher
terms are nonrenormalizable. Thus~(\ref{phil6}) amounts to the
idea to keep just all renormalizable terms in the effective action.
The same phenomenon --- that this is a good approximation ---
is observed for the 3d 3-state Potts model~\cite{StTs91},
and seems to be rather general.

\section{Conclusions}
We have studied the low-energy effective action for the scalar field
theory in three dimensions. We considered the 3d Ising model in the
symmetric phase in the scaling region. We have found that a
very good approximation is provided by
\begin{eqnarray}
  S_{eff} & = & \int d^3 x \left\{
  {1 \over 2} \partial_\mu \varphi \partial_\mu \varphi +
  V_{eff}(\varphi) \right\}, \\
  V_{eff}(\varphi) & = & {1 \over 2} m^2 \varphi^2 +
  m g_4 \varphi^4 + g_6 \varphi^6,  \label{eq33}
\end{eqnarray}
and estimated the values of dimensionless couplings:
\begin{equation}
  g_4 = 0.97 \pm 0.02, \quad g_6 = 2.05 \pm 0.15.
\end{equation}
(The errors are standard deviations). This effective action is
universal for the whole 3d Ising universality class, in the sense
that the only free parameter is the mass $m$. Its applicability
region is bounded by the requirement that $\varphi$ is not too
large. It works well at least for $\varphi$ such that
$m_{eff}(\varphi)=[V_{eff}''(\varphi)]^{1/2} \le 3m$, and maybe
further.

We observe a smooth approach to scaling in the 3d Ising model.
Our results for $g_4$ and $g_6$ are in good agreement with
the values available in the literature, but disagree
with~\cite{BeCo80}-\cite{BeBo93}. While, to our
knowledge, our result for $g_4$ is the most precise
among the available Monte Carlo estimates, and this is
the first Monte Carlo study to provide the definite
result on $g_6$, we consider the direct Monte Carlo
check that the approximation~(\ref{eq33}) works perfectly
as our main result.

This is in agreement with the results of Tetradis and Wetterich,
obtained within the effective average action
approach~\cite{TeWe93a,TeWe93b}, that $\phi^6$ is important, while
the higher terms in $V_{eff}$ provide only small corrections.

Our main tool was the study of the probability distribution
of the order parameter in the finite system with the periodic
boundary conditions, in the external field. We find that the
simple asymptotic formula~(\ref{preexp}) that connects the
probability distribution in a large, but finite system
with the infinite-volume effective potential works perfectly
in our case, when $L/\xi \ge 4$. The formula seems to be
rather general and useful also for other problems in the
theory of the order parameter distribution.

The Ising model in the symmetric phase is just the simplest
three-dimensional model. It would be interesting to extend
the approach developed here to its broken phase, to
weak first order transitions and to models that serve as
effective 3d theories for high-temperature phase transitions
in QCD and in the Higgs sector of the standard model.

\subsubsection*{Acknowledgements}
It is a pleasure to thank V.~Dohm, M.~E.~Fisher, M.~G\"ockeler,
J.~Jers\'ak, E.~Focht, C.~Frick, G.~M\"unster, S.~K.~Nechaev,
M.~A.~Stephanov, N.~Tetradis, J.~F.~Wheater and F.~Zimmerman for valuable
discussions and correspondence. I am deeply grateful
to Prof.~H.~A.~Kastrup, Prof.~J.~Jers\'ak and the
Institute of Theoretical Physics E of RWTH Aachen
for their kind hospitality, and to the Alexander von Humboldt
Foundation for the fellowship and other support.
This work was partially supported by the Russian Fund
for Fundamental Research Grant 93-02-3815.

\vfill
\newpage

\begin{table}
{\small
 \begin{center}
  \begin{tabular}{|l|l|l|l|l|l|l|}
   \hline
    $L^3$   & $14^3$  & $17^3$  & $22^3$  & $30^3$  & $38^3$  & $58^3$ \\
   \hline
    $\beta$ & 0.2115  & 0.2142  & 0.2167  & 0.2186  & 0.2195  & 0.22055  \\
    $J$     & 0.0     & 0.0     & 0.0     & 0.0     & 0.0     & 0.0      \\
            & 0.0043  & 0.0027  & 0.0014  & 0.00066 & 0.00036 & 0.00013  \\
            & 0.013   & 0.0081  & 0.0043  & 0.002   & 0.0011  & 0.00038  \\
    $N_{\rm conf}$ & $3 \times 700000$
               & $3 \times 720000$
               & $3 \times 700000$
               & $3 \times 400000$
               & $3 \times 500000$
               & $3 \times 100000$  \\
   \hline
    $Z^{-1}_{(J=0)}$ & 0.2294(2) & 0.2351(2) & 0.2407(2) & 0.2458(3) &
                    0.2493(3) & 0.2550(6)  \\
    $r$ & 0.01014(2) & 0.00690(3) & 0.00414(1) & 0.002263(6) &
         0.001468(5) & 0.000639(7) \\
    $u$ & 0.0163(2) & 0.01337(20) & 0.01062(11) & 0.00814(14) &
        0.00657(12) & 0.00451(17) \\
    $w$ & 0.0394(7) & 0.0388(8) & 0.0381(5) & 0.0360(10) &
         0.0361(12) & 0.0343(20) \\
   \hline
    $\sum \chi^2$  & 188 & 208 & 193 & 215 & 200 & 140 \\
    $N_{\rm bins}$ & 165 & 161 & 161 & 229 & 160 & 146 \\
   \hline
    $m$ & 0.2973(3) & 0.2423(5) & 0.1855(3) & 0.1357(2) &
          0.1085(2) & 0.0707(4) \\
    $g_4$ & 1.041(12) & 0.998(15) & 0.988(10) & 0.993(17) &
            0.974(18) & 0.98(4) \\
    $g_6$ & 3.26(6) & 2.99(6) & 2.73(4) & 2.42(7) & 2.33(8) & 2.07(12) \\
   \hline
  \end{tabular}
 \end{center}
}
\caption{The numerical results obtained by Monte Carlo.
$N_{\rm conf}$ is the number of configurations used.
$\sum \chi^2$ is the sum of $\chi^2$ for all histograms
(minimized by the fit); $N_{\rm bins}$ is the
total number of bins in the histograms. The numbers in the
parantheses are standard deviations of the last decimal
digits.} \label{table1}

\end{table}

\end{document}